\documentclass[twocolumn,showpacs,superscriptaddress]{revtex4-1}
\usepackage{graphicx}
\usepackage{units}
\usepackage{multirow}
\usepackage{bigstrut}
\usepackage{overpic}
\usepackage{array}
\usepackage{color}
\usepackage{amsmath}
\RequirePackage{lineno}
\newcommand{\ee}{e^+e^-}

\newcommand{\gevc}{\,\unit{GeV}/c}

\newcommand{\br}[1]{\mathcal{B}(#1)}

\begin{document}

\modulolinenumbers[2]

\setlength{\oddsidemargin}{-0.5cm} \addtolength{\topmargin}{15mm}

\title{\boldmath Measurement of the absolute branching fraction for
$\Lambda^+_{c}\to \Lambda e^+\nu_e$ }

\author{
  \begin{small}
    \begin{center}
      M.~Ablikim$^{1}$, M.~N.~Achasov$^{9,f}$, X.~C.~Ai$^{1}$,
      O.~Albayrak$^{5}$, M.~Albrecht$^{4}$, D.~J.~Ambrose$^{44}$,
      A.~Amoroso$^{49A,49C}$, F.~F.~An$^{1}$, Q.~An$^{46,a}$,
      J.~Z.~Bai$^{1}$
% ...
      \\
      \vspace{0.2cm}
      (BESIII Collaboration)\\
      \vspace{0.2cm} {\it
        $^{1}$ Institute of High Energy Physics, Beijing 100049, People's Republic of China\\
% ...
$^{4}$ Bochum Ruhr-University, D-44780 Bochum, Germany\\
        $^{5}$ Carnegie Mellon University, Pittsburgh, Pennsylvania 15213, USA\\
% ...
$^{9}$ G.I. Budker Institute of Nuclear Physics SB RAS (BINP), Novosibirsk 630090, Russia\\
% ...
$^{44}$ University of Rochester, Rochester, New York 14627, USA\\
% ...
$^{46}$ University of Science and Technology of China, Hefei 230026, People's Republic of China\\
% ...
$^{49}$ (A)University of Turin, I-10125, Turin, Italy; (B)University of Eastern Piedmont, I-15121, Alessandria, Italy; (C)INFN, I-10125, Turin, Italy\\
% ...
$^{a}$ Also at State Key Laboratory of Particle Detection and Electronics, Beijing 100049, Hefei 230026, People's Republic of China\\
% ...
$^{f}$ Also at the Novosibirsk State University, Novosibirsk, 630090, Russia\\
% ...
      }\end{center}
    \vspace{0.4cm}
  \end{small}
}

\begin{abstract}
We report the first absolute measurement of the branching fraction
of $\Lambda^+_{c}\rightarrow \Lambda e^+\nu_e$. This measurement is
based on 567 pb$^{-1}$ of $\ee$ annihilation data produced at
$\sqrt{s}=4.599$ GeV, which is just above the
$\Lambda^+_c\bar{\Lambda}^-_c$ threshold.  The data were collected
with the BESIII detector at the BEPCII storage rings. The branching
fraction is determined to be $\mathcal B({\Lambda^+_c\rightarrow
\Lambda e^+\nu_e})=(3.63\pm0.38({\rm stat})\pm0.20({\rm syst}))\%$,
representing a more than twofold improvement in precision upon
previously published results. As the branching fraction for
$\Lambda^+_{c}\rightarrow \Lambda e^+\nu_e$ is the benchmark for
those of other $\Lambda^+_c$ semileptonic channels, our result
provides a unique test of different theoretical models, which is the
most stringent to date.
\end{abstract}


\maketitle


Semileptonic (SL) decays of the lightest charmed baryon,
$\Lambda_c^+$, provide a stringent test for non-perturbative aspects
of the theory of strong interaction.  In particular, the decay rate
of the most copious SL decay mode, $\Lambda_c^+ \to \Lambda e^+
\nu_e$, serves as a benchmark for all other $\Lambda_c^+$ SL decay
rates. The $\Lambda^+_c\rightarrow \Lambda e^+\nu_e$ decay is
dominated by the Cabibbo-favored transition $c\rightarrow s
l^+\nu_l$, which occurs, to a good approximation, independently of
the spin-zero spectator $ud$ diquark. This leads to a simpler
theoretical description and greater predictive power in modeling the
SL decays of the charmed baryons than the case for
mesons~\cite{Richman:1995wm}. However, model development for
semileptonic decays of charmed mesons is much more advanced because
of the availability of experimental data with precision better than
5\%~\cite{pdg2014}. An experimental study of $\Lambda^+_c\rightarrow
\Lambda e^+\nu_e$ is therefore desirable in order to test different
models in the charm baryon sector~\cite{charg}.

Since the first observation of the $\Lambda^+_c$ baryon in $e^+e^-$
annihilations at the Mark II experiment~\cite{prl44_10} in 1979, much
theoretical effort has been applied towards the study of its
SL decay properties. However, predictions of the branching fraction
(BF) $\br{\Lambda_c^+\rightarrow \Lambda e^+\nu_e}$ in different
theoretical models vary in a wide range from 1.4\% to
9.2\%~\cite{prd40_2955,prd40_2944,zpc51_607,zpc52_149,prd43_2939,prd45_3266,prd53_1457,plb_431_173,prd60_034009,prc72_032005,prd80_074011},
depending on the choice of various $\Lambda_c^+$ wave function
models and the nature of decay dynamics. In addition, theoretical calculations prove to be quite
challenging for lattice quantum chromodynamics (LQCD) due to the
complexity of form factors, which describes the hadronic part of the decay dynamics in $\Lambda^+_c\rightarrow \Lambda
e^+\nu_e$~\cite{prd86_014017}. Thus, an accurate measurement of
$\br{\Lambda_c^+\rightarrow \Lambda e^+\nu_e}$
is a key ingredient in calibrating LQCD calculations, which, in turn,
will play an important role in understanding  different
$\Lambda_c^+$ SL decays.

So far, experimental information for $\br{\Lambda_c^+\rightarrow
\Lambda e^+\nu_e}$ has come only from the ARGUS~\cite{plb269_234}
and CLEO~\cite{plb323_219} experiments in the 1990s. They measured
the product cross section $\sigma(e^+e^-\rightarrow
\Lambda_c^+X)\cdot\br{\Lambda_c^+\rightarrow \Lambda
e^+\nu_e}$ at $B\bar{B}$ threshold energies. Combined with the
measured $\br{\Lambda_c^+\rightarrow
pK^-\pi^+}=(5.0\pm1.3)\%$ and the $\Lambda^+_c$ lifetime, they
evaluated $\br{\Lambda^+_c\rightarrow \Lambda
e^+\nu_e}=(2.1\pm0.6)\%$~\cite{pdg2014}. Therefore, this is not a
direct determination of $\mathcal{B}(\Lambda_c^+\rightarrow \Lambda
e^+\nu_e)$. In this Letter, we report the first absolute measurement
of $\mathcal{B}(\Lambda_c^+\rightarrow \Lambda e^+\nu_e)$ by
analyzing $567$ pb$^{-1}$~\cite{lum} of data accumulated at
$\sqrt{s} = 4.599$ GeV with the BESIII detector at the BEPCII
collider. This is the largest $\Lambda^+_c$ data sample near the
$\Lambda^+_c\bar{\Lambda}^-_c$ threshold, where the $\Lambda^+_c$ is
always produced in association with a $\bar{\Lambda}^-_c$ baryon.
Hence, $\br{\Lambda^+_{c}\rightarrow \Lambda e^+\nu_e}$ can be
accessed by measuring the relative probability of finding the SL decay
when the $\bar{\Lambda}^-_c$ is reconstructed in a number of prolific decay
channels. This will provide a clean and straightforward BF
measurement without requiring knowledge of the total number of
$\Lambda^+_c\bar{\Lambda}^-_c$ events produced.

BESIII~\cite{Ablikim:2009aa} is a cylindrical spectrometer,
which is composed of a Helium-gas based main drift chamber (MDC), a
plastic scintillator time-of-flight (TOF) system, a CsI~(Tl)
electromagnetic calorimeter (EMC), a superconducting solenoid
providing a 1.0\,T magnetic field and a muon counter. The  charged
particle momentum resolution is 0.5\% at a transverse momentum of
1\,$\gevc$ and the photon energy resolution is 2.5\% at 1\,GeV.
Particle identification (PID) system combines the ionization energy
loss ($dE/dx$) in MDC, the TOF and EMC information to identify
particle types. More details about the design and performance of the
detector are given in Ref.~\cite{Ablikim:2009aa}.

A GEANT4-based~\cite{geant4} Monte Carlo (MC) simulation package,
which includes the geometric description of the detector and the
detector response, is used to determine the detection efficiency and
to estimate the potential backgrounds. Signal MC samples of a
$\Lambda_c$ baryon decaying only to $\Lambda e\nu_e$ together with a
$\bar{\Lambda}_c$ decaying only to the studied tag modes are generated
by the MC generator KKMC~\cite{kkmc} using
EVTGEN~\cite{nima462_152}, with initial-state radiation (ISR)
effects~\cite{SJNP41_466} and final-state radiation
effects~\cite{plb303_163} included.
For the simulation of the decay $\Lambda_c^+\rightarrow \Lambda
e^+\nu_e$, we use the form factor predictions obtained using Heavy Quark
Effective Theory and QCD sum rules of Ref.~\cite{prd60_034009}.
To study backgrounds, `inclusive' MC samples consisting of
$\Lambda_c^+\bar{\Lambda}_c^-$ events, $D_{(s)}$ production, ISR
return to the charmonium(-like) $\psi$ states at lower masses and continuum processes are
generated. All decay modes of the $\Lambda_c$, $\psi$ and $D_{(s)}$
as specified in the Particle Data Group (PDG)\cite{pdg2014} are
simulated by the MC generator. The unknown decays of the $\psi$ states are
generated with LUNDCHARM~\cite{lundcharm}.

The technique for this analysis, which was first applied by the
Mark~III Collaboration~\cite{prl62_1821} at SPEAR, relies on the
purity and kinematics of the $\Lambda_c^+\bar{\Lambda}^-_c$ baryon pairs
produced at $\sqrt{s}=4.599$ GeV. First, we select
a data sample of $\bar{\Lambda}^-_c$ baryons by reconstructing exclusive
hadronic decays; we call this the single tag (ST) sample. Then, we search
for $\Lambda_c^+\rightarrow \Lambda e^+\nu_e$ in the system recoiling against the ST
$\bar{\Lambda}^-_c$ baryons. The ST $\bar{\Lambda}^-_c$ baryons are
reconstructed using eleven hadronic decay modes:
$\bar{\Lambda}^-_c\rightarrow \bar{p} K^0_S$, $\bar{p} K^+\pi^-$,
$\bar{p}K^0_S\pi^0$, $\bar{p} K^+\pi^-\pi^0$, $\bar{p}
K^0_S\pi^+\pi^-$, $\bar{\Lambda}\pi^-$, $\bar{\Lambda}\pi^-\pi^0$,
$\bar{\Lambda}\pi^-\pi^+\pi^-$, $\bar{\Sigma}^0\pi^-$,
$\bar{\Sigma}^-\pi^0$ and $\bar{\Sigma}^-\pi^+\pi^-$, where the
intermediate particles $K^0_S$, $\bar{\Lambda}$, $\bar{\Sigma}^0$,
$\bar{\Sigma}^-$ and $\pi^0$ are reconstructed by their decays into
$K^0_S\rightarrow \pi^+\pi^-$, $\bar{\Lambda}\rightarrow
\bar{p}\pi^+$, $\bar{\Sigma}^0\rightarrow \gamma\bar{\Lambda}$ with
$\bar{\Lambda}\rightarrow \bar{p}\pi^+$, $\bar{\Sigma}^-\rightarrow
\bar{p}\pi^0$ and $\pi^0\rightarrow \gamma\gamma$, respectively.

Charged tracks are required to have polar angles within
$|\cos\theta|<0.93$, where $\theta$ is the polar angle of the
charged track with respect to the beam direction. Their distances of
closest approach to the interaction point (IP) are required to be
less than 10\,cm along the beam direction and less than 1\,cm in the
perpendicular plane. Tracks originating from $K^0_S$ and $\Lambda$
decays are not subjected to these distance requirements. To
discriminate pions from kaons, the $dE/dx$ and TOF information are
used to obtain probabilities for the pion ($\mathcal{L}_{\pi}$) and kaon
($\mathcal{L}_K$) hypotheses. Pions and kaons are identified by
$\mathcal{L}_{\pi} > \mathcal{L}_{K}$ and $\mathcal{L}_{K} >
\mathcal{L}_{\pi}$, respectively. For proton identification,
information from $dE/dx$, TOF, and EMC are combined to calculate the PID
probability $\mathcal{L'}$, and a charged track satisfying
$\mathcal{L'}_p> \mathcal{L'}_{\pi}$ and $\mathcal{L'}_{p} >
\mathcal{L'}_{K}$ is identified as a proton candidate.

Photon candidates are reconstructed from isolated clusters in the
EMC in the regions $|\cos\theta| \le 0.80$ (barrel) and $0.86 \le |\cos\theta|
  \le 0.92$ (end cap). The deposited energy of a neutral cluster is required to be larger than 25 (50) MeV in
barrel (end cap) region, and the angle between the photon candidate
and the nearest charged track must be larger than 10$^\circ$.
To suppress electronic noise and energy deposits unrelated to the
events, the difference between the EMC time and the event start time is required
to be within (0, 700)~ns. To reconstruct $\pi^0$ candidates,
the invariant mass of the accepted photon pairs is required to be within
$(0.110,~0.155)$~GeV$/c^2$. A kinematic fit is implemented to
constrain the $\gamma\gamma$ invariant mass to the $\pi^0$ nominal
mass~\cite{pdg2014}, and the $\chi^2$ of the kinematic fit is
required to be less than 20. The fitted momenta of the $\pi^0$ are
used in the further analysis.

To reconstruct $K^0_S$ and $\bar{\Lambda}$, a secondary vertex fit is
applied, and the decay length is required to be larger than zero. The
invariant masses $M(\pi^+\pi^-)$, $M(\bar{p}\pi^+)$,
$M(\gamma\bar{\Lambda})$ and $M(\bar{p}\pi^0)$ are required to be
within $(0.485,~0.510)$~GeV/$c^2$, $(1.110,~1.121)$~GeV/$c^2$,
$(1.179,~1.205)$~GeV/$c^2$ and $(1.173,~1.200)$~GeV/$c^2$ to select
candidates for $K^0_S$, $\bar{\Lambda}$, $\bar{\Sigma}^0$ and
$\bar{\Sigma}^-$, respectively.

For the ST mode of $\bar{p}K^0_S\pi^0$, $\bar{\Lambda}$ and
$\bar{\Sigma}^-$ backgrounds are rejected by vetoing any events with
$M(\bar{p}\pi^+)$ and $M(\bar{p}\pi^0)$ inside the regions $(1.105,
1.125)$~GeV/$c^2$ and $(1.173, 1.200)$~GeV/$c^2$, respectively. For
the ST modes of $\bar{\Lambda}\pi^+\pi^-\pi^-$ and
$\bar{\Sigma}^-\pi^+\pi^-$, $K^0_S$ backgrounds are suppressed by
requiring $M(\pi^+\pi^-)$ outside of $(0.480, 0.520)$~GeV/$c^2$,
while $\Lambda$ backgrounds are removed from decays to $\bar{p}K^0_S\pi^+\pi^-$
and $\bar{\Sigma}^-\pi^+\pi^-$ by requiring
$M(\bar{p}\pi^+)$ to be outside of $(1.105, 1.125)$~GeV/$c^2$.

\begin{table}[tp!]
\caption{ $\Delta E$ requirements and  ST yields
$N_{\bar{\Lambda}_c^-}$ in data.}
\begin{center}
\begin{tabular}
{llc} \hline \hline Mode~&~~~$\Delta E$(GeV)~~~&$N_{\bar{\Lambda}_c^-}$ \\
\hline
 $\bar{p} K^0_S$                & [$-$0.025, 0.028] &   $1066\pm33$  \\
 $\bar{p} K^+\pi^-$             & [$-$0.019, 0.023] &   $5692\pm88$  \\
 $\bar{p}K^0_S\pi^0$            & [$-$0.035, 0.049] &  ~~$593\pm41$  \\
 $\bar{p} K^+\pi^-\pi^0$        & [$-$0.044, 0.052] &   $1547\pm61$  \\
 $\bar{p} K^0_S\pi^+\pi^-$      & [$-$0.029, 0.032] &  ~~$516\pm34$  \\
 $\bar{\Lambda}\pi^-$           & [$-$0.033, 0.035] &  ~~$593\pm25$  \\
 $\bar{\Lambda}\pi^-\pi^0$      & [$-$0.037, 0.052] &   $1864\pm56$  \\
 $\bar{\Lambda}\pi^-\pi^+\pi^-$ & [$-$0.028, 0.030] &  ~~$674\pm36$  \\
 $\bar{\Sigma}^0\pi^-$          & [$-$0.029, 0.032] &  ~~$532\pm30$  \\
 $\bar{\Sigma}^-\pi^0$          & [$-$0.038, 0.062] &  ~~$329\pm28$  \\
 $\bar{\Sigma}^-\pi^+\pi^-$     & [$-$0.049, 0.054] &   $1009\pm57$  \\
\hline \hline
\end{tabular}
\label{tab:deltaE_1}
\end{center}
\end{table}

The ST $\bar{\Lambda}^-_c$ signals are identified using the beam
constrained mass, $M_{\rm BC}=\sqrt{E^2_{\rm
beam}-|\overrightarrow{p}_{\bar{\Lambda}^-_c}|^2}$, where $E_{\rm
beam}$ is the beam energy and
$\overrightarrow{p}_{\bar{\Lambda}^-_c}$ is the momentum of the
$\bar{\Lambda}^-_c$ candidate. To improve the signal purity, the
energy difference $\Delta E=E_{\rm beam}-E_{\bar{\Lambda}^-_c}$ for
each candidate is required to be within approximately
$\pm3\sigma_{\Delta E}$ around the $\Delta E$ peak, where
$\sigma_{\Delta E}$ is the $\Delta E$ resolution and
$E_{\bar{\Lambda}^-_c}$ is the reconstructed $\bar{\Lambda}^-_c$
energy. The explicit $\Delta E$ requirements for different modes are
listed in Table~\ref{tab:deltaE_1}.

\begin{figure}[tp!]
\begin{center}
\includegraphics[width=\linewidth]{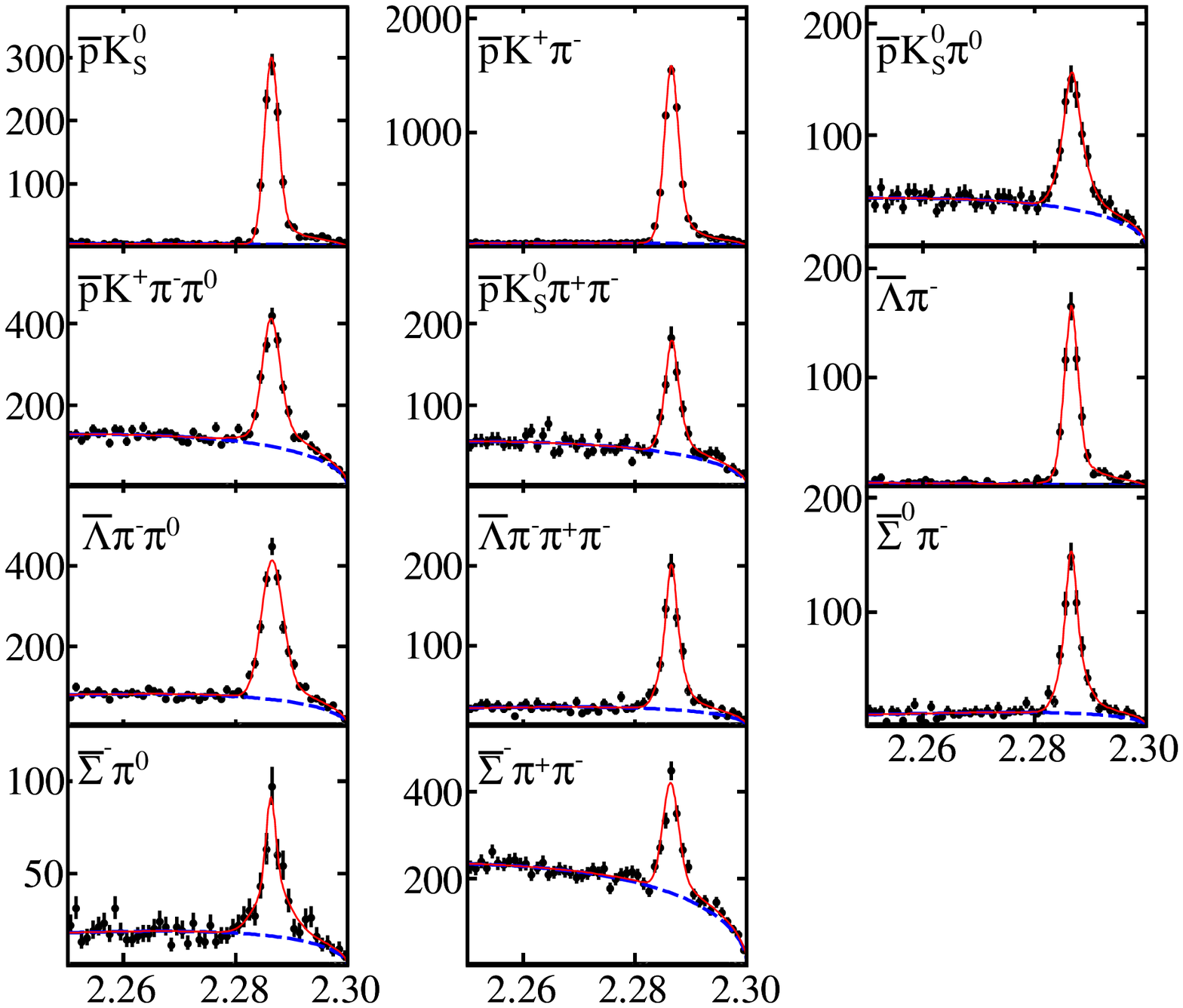}
   \put(-248,82){\rotatebox{90}{\normalsize Events/0.001 GeV/$c^2$}}
   \put(-140,5){\normalsize $M_{\rm BC}$ (GeV$/c^2$) }
\caption{Fits to the $M_{\rm BC}$ distributions for different ST modes.
The points with error bars are data, the (red) solid curves show the total
fits and the (blue) dashed curves are the background shapes. }
\label{fig:tag_lambdac}
\end{center}
\end{figure}

The $M_{\rm BC}$ distributions for the eleven $\bar{\Lambda}^-_c$ ST
modes are shown in Fig.~\ref{fig:tag_lambdac}. We perform unbinned
maximum likelihood fits to the spectra, where we use the MC
simulated signal shape convoluted with a double-Gaussian resolution
function to represent the signal shape and an ARGUS
function~\cite{plb241_278} to describe the background shape. The
signal yield is estimated in the  mass
region $(2.280, 2.296)$~GeV/$c^2$. Peaking backgrounds are evaluated to be about 0.3\%,
according to MC simulations. These backgrounds are subtracted
from the fitted number of the singly tagged $\bar{\Lambda}_c^-$
events. The numbers of background-subtracted signal events are used as the ST
yields, as listed in Table~\ref{tab:deltaE_1}. Finally, we obtain
the total ST yield summed over all 11 modes to be $N^{\rm
tot}_{\bar{\Lambda}^-_c}=14415\pm159$.

Candidate events for $\Lambda^+_c\rightarrow \Lambda e^+\nu_e$ are
selected from the remaining tracks recoiling against the ST
$\bar{\Lambda}^-_c$ candidates. To select the $\Lambda$, the same
criteria as those used in the ST selection are applied. We further
identify a charged track as an $e^+$ by requiring the probabilities
calculated with the $dE/dx$, TOF and EMC satisfying the criteria
$\mathcal{L}'_{e} > 0.001$ and
$\mathcal{L}'_e/(\mathcal{L}'_e+\mathcal{L}'_{\pi}+\mathcal{L}'_K)>0.8$.
Its energy loss due to
bremsstrahlung photon(s) is partially recovered by adding the
showers that are within a 5$^{\circ}$ cone about the positron momentum.
As the neutrino is not detected, we employ the kinematic variable
$$U_{\rm miss}=E_{\rm miss}-c|\vec{p}_{\rm miss}|$$
to obtain information on the neutrino, where $E_{\rm
miss}$ and $\vec{p}_{\rm miss}$ are the missing energy and momentum
carried by the neutrino, respectively. They are calculated by
$E_{\rm miss}=E_{\rm beam}-E_{\Lambda}-E_{e^+}$ and $\vec{p}_{\rm
miss}=\vec{p}_{\Lambda_c^+}-\vec{p}_{\Lambda}-\vec{p}_{e^+}$, where
$\vec{p}_{\Lambda_c^+}$ is the momentum of $\Lambda_c^+$ baryon, $E_{\rm
\Lambda}$($\vec{p}_{\Lambda}$) and $E_{e^+}$ ($\vec{p}_{e^+}$) are
the energies (momenta) of the $\Lambda$ and the positron,
respectively. Here, the momentum $\vec{p}_{\Lambda_c^+}$ is given by
$\vec{p}_{\Lambda_c^+}=-\hat{p}_{\rm tag}\sqrt{E_{\rm
beam}^2-m^2_{\bar{\Lambda}^-_c}}$, where $\hat{p}_{\rm tag}$ is the
direction of the momentum of the ST $\bar{\Lambda}^-_c$ and
$m_{\bar{\Lambda}^-_c}$ is the nominal $\bar{\Lambda}^-_c$
mass~\cite{pdg2014}. For signal events, $U_{\rm miss}$
is expected to peak around zero.

\begin{figure}[tp!]
\begin{center}
   \begin{minipage}[t]{4.2cm}
   \includegraphics[height=3.7cm,width=4.4cm]{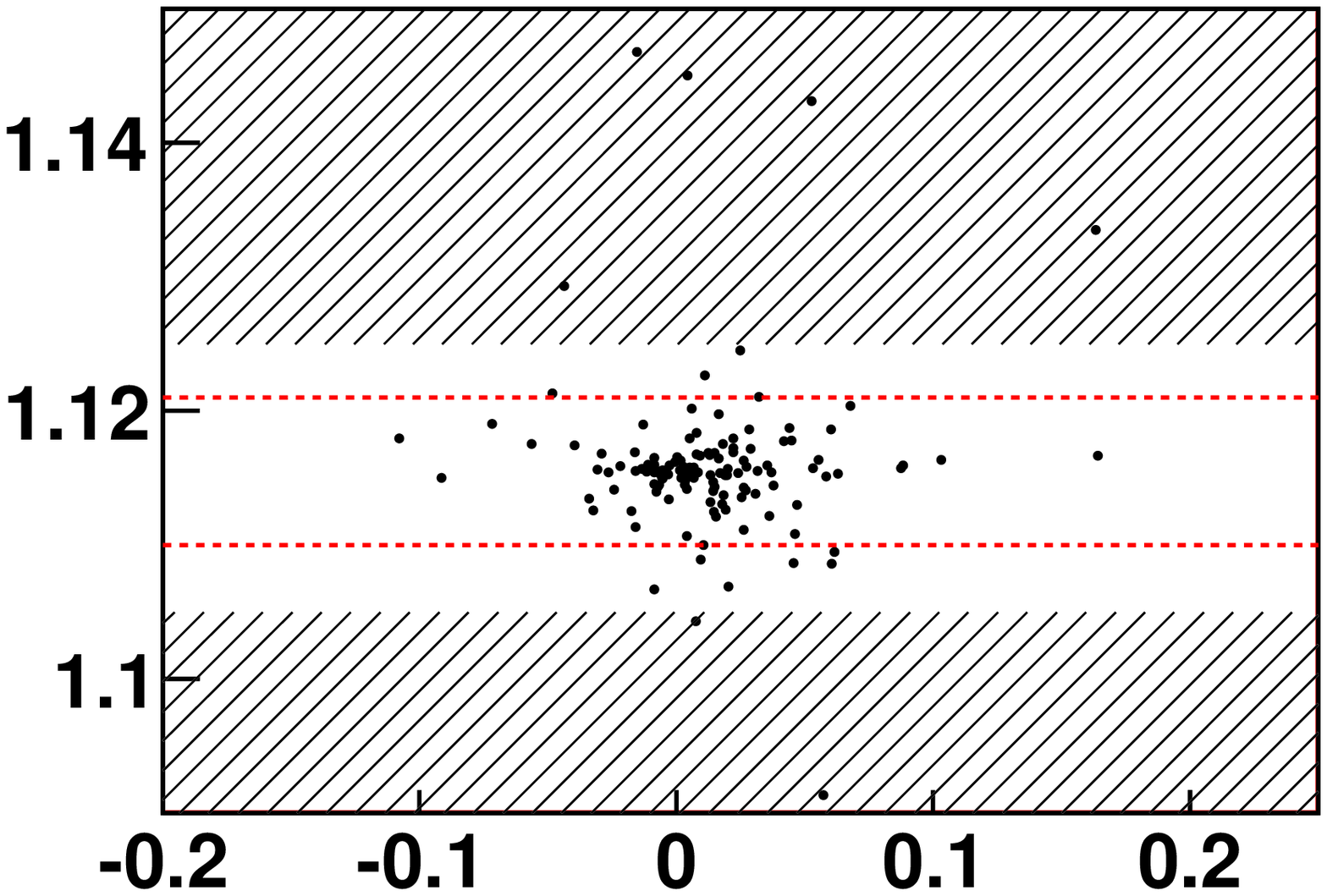}
   \put(-35,85){\bf \scriptsize (a)}
   \put(-130,22){\rotatebox{90}{\small $M_{p\pi^-}$ (GeV$/c^2$)}}
   \put(-80,-5){\normalsize $U_{\rm miss}$ (GeV) }
   \end{minipage}
   \begin{minipage}[t]{4.2cm}
   \includegraphics[height=3.7cm,width=4.4cm]{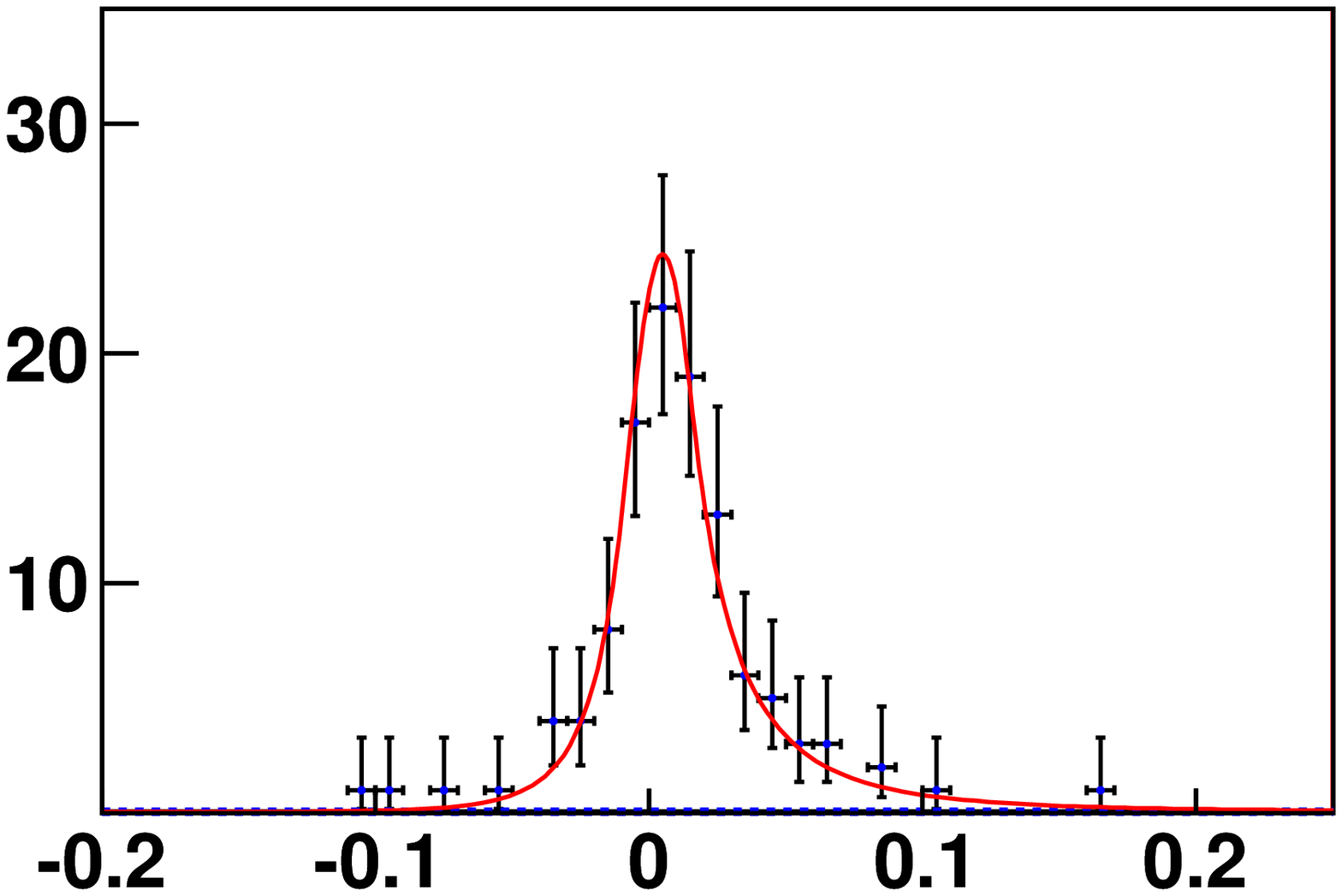}
   \put(-35,85){\bf \scriptsize (b)}
   \put(-130,16){\rotatebox{90}{\footnotesize Events/0.010 GeV}}
   \put(-80,-5){\normalsize $U_{\rm miss}$ (GeV) }
   \end{minipage}
   \caption{ (a) Scatter plot of $M_{p\pi^-}$ versus $U_{\rm miss}$ for the $\Lambda_c^+\rightarrow \Lambda e^+\nu_e$
   candidates. The area between the dashed lines denotes the $\Lambda$ signal region and the hatched areas indicate the $\Lambda$ sideband regions. (b) Fit to the $U_{\rm miss}$
distribution within the $\Lambda$ signal region. The points with
error bars are data, the (red) solid curve shows the total fit and
the (blue) dashed curve is the background shape.}
\label{fig:umiss_data_sig}
\end{center}
\end{figure}

Figure~\ref{fig:umiss_data_sig}(a) shows a scatter plot of
$M_{p\pi^-}$ versus $U_{\rm miss}$ for the $\Lambda^+_c\to \Lambda
e^+\nu_e$ candidates in data. Most of the events are located around
the intersection of the $\Lambda$ and $\Lambda e^+\nu_e$ signal
regions. Requiring $M_{p\pi^-}$ to be within the $\Lambda$ signal
region, we project the scatter plot onto the $U_{\rm miss}$ axis, as
shown in Fig.~\ref{fig:umiss_data_sig}(b). The $U_{\rm miss}$
distribution is fitted with a signal function $f$ plus a polynomial
function to describe the background. The signal function
$f$~\cite{prd79_052010} consists of a Gaussian function to model the
core of the $U_{\rm miss}$ distribution and two power law tails to
account for the effects of initial and final state radiation:
\begin{equation}
f(U_{\rm miss})=\left\{
\begin{array}{rcl}
p_1(\frac{n_1}{\alpha_1}-\alpha_1+t)^{-n_1},      &      & t>\alpha_1\\
e^{-t^2/2},~~~~~~~~~~~~~~~~     &      & -\alpha_2<t<\alpha_1 \\
p_2(\frac{n_2}{\alpha_2}-\alpha_2-t)^{-n_2},       &      &
t<-\alpha_2
\end{array} \right.
\label{eq:fun_umiss}
\end{equation}
where $t=(U_{\rm miss}-U_{\rm mean})/\sigma_{U_{\rm miss}}$, $U_{\rm
mean}$ and $\sigma_{U_{\rm miss}}$ are the mean value and resolution
of the Gaussian function, respectively,
$p_1\equiv(n_1/\alpha_1)^{n_1}e^{-\alpha^2_1/2}$ and
$p_2\equiv(n_2/\alpha_2)^{n_2}e^{-\alpha^2_2/2}$. The parameters
$\alpha_1$, $\alpha_2$, $n_1$ and $n_2$ are fixed to the values
obtained in the signal MC simulations. From the fit, we obtain the
number of SL signals to be $109.4\pm10.9$.

The backgrounds in $\Lambda_c^+\rightarrow \Lambda e^+\nu_{e}$ arise
mostly from misreconstructed SL decays with correctly reconstructed
tags. There are two types of peaking backgrounds. The first comes
from non-$\Lambda$ SL decays, which are studied using data in the
$\Lambda$ sideband in Fig.~\ref{fig:umiss_data_sig}. We obtain the
number of events of the first type of backgrounds to be $1.4\pm0.8$,
after scaling to the $\Lambda$ signal region. The second peaking background arises from
$\Lambda_c^+\rightarrow\Lambda\mu^+\nu_{\mu}$ and some hadronic
decays, such as $\Lambda_c^+\rightarrow\Lambda\pi^+\pi^0$,
$\Lambda\pi^+$ and $\Sigma^0\pi^+$. Based on MC simulations, we
determine the number of background events of the second type to
be $4.5\pm0.5$. After subtracting these background events, we determine
the net number of $\Lambda_c^+\rightarrow \Lambda e^+\nu_{e}$ to be
$N_{\rm semi}=103.5\pm10.9$, where the uncertainty is statistical.

The absolute BF for $\Lambda_c^+\rightarrow \Lambda e^+\nu_e$ is
determined by
\begin{equation} \mathcal{B}(\Lambda_c^+\rightarrow \Lambda e^+\nu_e)=\frac{N_{\rm semi}}{N^{\rm tot}_{\bar{\Lambda}_c^-}\times\varepsilon_{\rm
semi}\times\mathcal{B}(\Lambda\rightarrow p\pi^-)},
\label{eq:branch}
\end{equation}
where $\varepsilon_{\rm semi}=(30.92\pm0.26)\%$, which does not
include the BF for $\Lambda\rightarrow p\pi^-$, is the overall
efficiency for detecting the $\Lambda_c^+\rightarrow \Lambda
e^+\nu_e$ decay in ST events, weighted by the ST
yields of data for each tag.
Inserting the values of $N_{\rm semi}$, $N^{\rm
tot}_{\bar{\Lambda}^-_c}$, $\epsilon_{\rm semi}$ and
$\mathcal{B}(\Lambda\rightarrow p\pi^-)$~\cite{pdg2014} in
Eq.~(\ref{eq:branch}), we get $\mathcal B({\Lambda^+_c\rightarrow
\Lambda e^+\nu_e})=(3.63\pm0.38\pm0.20)\%$, where the first error is
statistical and the second systematic.

The systematic error is
mainly due to the uncertainty in the efficiency of $\Lambda$ reconstruction (2.5\%),
which is studied with $\chi_{cJ}\rightarrow
\Lambda\bar{\Lambda}\pi^+\pi^-$, and the simulation of the SL signal
model (4.5\%), estimated by changing the default parameterization of
form factor function to other parameters in Refs.~\cite{prd60_034009,Hinson:2004pj}
and by taking into account the $q^2$ dependence observed in data.
Other relevant issues include the following uncertainties: the
electron tracking (1.0\%) and the electron PID (1.0\%) which is
studied with $e^+e^-\rightarrow (\gamma)e^+e^-$, the fit to the
$U_{\rm miss}$ distribution (0.8\%) estimated by using alternative
signal shapes, the quoted BF for $\Lambda\rightarrow p\pi^-$
(0.8\%), the MC statistics (0.8\%), the background subtraction
(0.5\%), the $N_{\bar{\Lambda}_c^-}$ (1.0\%) evaluated by using
alternative signal shapes in the fits to the $M_{\rm BC}$ spectra.
The total systematic error is estimated to be 5.6\% by adding all
these uncertainties in quadrature.

In summary, we report the first absolute measurement of the BF for
$\Lambda^+_c\rightarrow \Lambda e^+\nu_e$, $\mathcal
B({\Lambda^+_c\rightarrow \Lambda
e^+\nu_e})=(3.63\pm0.38\pm0.20)\%$, based on 567~pb$^{-1}$ data
taken at $\sqrt{s}=4.599$~GeV. This work improves the precision of
the world average value more than twofold. As the theoretical
predictions on this rate vary in a large range of
$1.4-9.2\%$~\cite{prd40_2955,prd40_2944,zpc51_607,zpc52_149,prd43_2939,prd45_3266,prd53_1457,plb_431_173,prd60_034009,prc72_032005,prd80_074011},
our result thus provide a stringent test on these non-perturbative
models. At a confidence level of $95\%$, this measurement disfavors
the predictions in
Refs.~\cite{prd40_2955,prd40_2944,zpc52_149,prd43_2939,prd45_3266}.

The BESIII collaboration thanks the staff of BEPCII and the IHEP
computing center for their strong support. This work is supported in
part by National Key Basic Research Program of China under Contract
No. 2015CB856700; National Natural Science Foundation of China (NSFC)
under Contracts Nos. 11125525, 11235011, 11275266, 11322544, 11322544, 11335008,
11425524; the Chinese Academy of Sciences (CAS) Large-Scale Scientific
Facility Program; the CAS Center for Excellence in Particle Physics
(CCEPP); the Collaborative Innovation Center for Particles and
Interactions (CICPI); Joint Large-Scale Scientific Facility Funds of
the NSFC and CAS under Contracts Nos. 11179007, U1232201, U1332201;
CAS under Contracts Nos. KJCX2-YW-N29, KJCX2-YW-N45; 100 Talents
Program of CAS; National 1000 Talents Program of China; INPAC and
Shanghai Key Laboratory for Particle Physics and Cosmology; German
Research Foundation DFG under Contract No. Collaborative Research
Center CRC-1044; Istituto Nazionale di Fisica Nucleare, Italy;
Koninklijke Nederlandse Akademie van Wetenschappen (KNAW) under
Contract No. 530-4CDP03; Ministry of Development of Turkey under
Contract No. DPT2006K-120470; Russian Foundation for Basic Research
under Contract No. 14-07-91152; The Swedish Resarch Council;
U. S. Department of Energy under Contracts Nos. DE-FG02-04ER41291,
DE-FG02-05ER41374, DE-SC0012069, DESC0010118; U.S. National Science
Foundation; University of Groningen (RuG) and the Helmholtzzentrum
fuer Schwerionenforschung GmbH (GSI), Darmstadt; WCU Program of
National Research Foundation of Korea under Contract
No. R32-2008-000-10155-0.



\begin{thebibliography}{99}
\bibitem{charg} Throughout the text,
the inclusion of charge conjugated processes is always implied.
\end{thebibliography}
\end{document}